\documentclass[prl,twocolumn, aps,amssymb,footinbib,showpacs]{revtex4-1}

\usepackage{amsmath, amssymb, braket, dsfont, times, tensor}
\usepackage[mathscr]{euscript}
\usepackage{graphicx}
\usepackage{subfigure}		
\usepackage{epstopdf, xcolor}
\usepackage[breaklinks]{hyperref}

\begin{document}
\title{Preparation of Low Entropy Correlated Many-body States via Conformal Cooling Quenches }
\author{Michael P. Zaletel$^{1}$, Adam M. Kaufman$^{2}$, Dan M. Stamper-Kurn$^{1}$,  Norman Y. Yao$^{1}$}
\affiliation{$^{1}$Department of Physics, University of California Berkeley, Berkeley, CA 94720, U.S.A.}
\affiliation{$^{2}$JILA, University of Colorado and National Institute of Standards and Technology,
and Department of Physics, University of Colorado, Boulder, Colorado 80309, U.S.A.}

\date{\today}							
\begin{abstract}
We propose and analyze a method for preparing low-entropy many-body states in isolated quantum optical systems of atoms, ions and molecules. 
Our approach is based upon shifting entropy between different regions of a system by spatially modulating the magnitude of the effective Hamiltonian. 
We conduct two case studies, on a topological spin chain and the spinful fermionic Hubbard model, focusing on the key question: can a ``conformal cooling quench'' remove sufficient entropy within experimentally accessible  timescales? 
Finite temperature, time-dependent matrix product state calculations reveal that even moderately sized ``bath'' regions can remove enough energy and entropy density to expose coherent low temperature physics. 
The protocol is particularly natural in systems with long-range interactions such lattice-trapped polar molecules and Rydberg-excited atoms where the magnitude of the Hamiltonian scales directly with the interparticle spacing. 
To this end, we propose simple, near-term implementations of conformal cooling quenches in systems of atoms or molecules, where signatures of low-temperature phases may be observed.

\end{abstract}

\maketitle

Ultracold quantum gases have reached the extraordinary realm of sub-nanokelvin temperatures \cite{leanhardt2003cooling,kovachy2015matter}, revealing, along the way, phenomena ranging from Bose-Einstein condensation and Cooper-paired superfluidity to  Mott insulators and localization \cite{anderson1995dilute,davis1995bose,demarco1999onset,chin2004observation,schreiber2015observation}. This scientific impact owes, in part, to a flexible array of cooling techniques that can effectively quench the kinetic energy of atomic systems; indeed, the laser cooling of atomic registers in optical tweezers has enabled the observation of few-particle quantum interference and entanglement \cite{kaufman2012cooling,kaufman2014two}, while the evaporative cooling of Bose gases has realized temperatures nearly two orders of magnitude smaller than that required for condensation \cite{olf2015thermometry}.  

Nevertheless, these temperatures are still too high to emulate a number of more exotic- and delicate- quantum phases including antiferromagnetic spin liquids, fractional Chern insulators and high-temperature superconductors \cite{stamper2009viewpoint,campbell2011ultracold,yao2013realizing}. 
The figure of merit for observing such physics is not the absolute temperature, but rather the dimensionless entropy density in units of $k_B$~\cite{quantummontecarlo}.
Reaching ultra-low entropy densities remains a major challenge for many-body quantum simulations despite the multitude of kinetic cooling techniques. This challenge is particularly acute for gases in deep optical lattice potentials, for which transport, and thus evaporative cooling, is slowed \cite{hung2010slow}. Moreover, in lattice systems representing models of quantum magnetism, the entropy resides primarily in spin, rather than motional, degrees of freedom \cite{chu2002cold}.  Expelling such entropy through evaporative cooling requires the conversion of spin excitations to kinetic excitations, a process that is typically inefficient \cite{hart2015observation,parsons2016site,cheuk2016observation}.

To access low-entropy phases of matter, two broad approaches have been proposed toward overcoming this challenge. 
The first is adiabatic preparation, where one initializes a low entropy  state and changes the Hamiltonian gradually until the desired many-body state is reached \cite{sorensen2010adiabatic,barkeshli2015continuous, Chiu2018}. 
However, the final entropy density is bounded from below by the initial entropy density, and experimental constraints or  phase transitions may preclude a suitable adiabat.
The second approach is to `shift entropy elsewhere' \cite{stamper2009viewpoint,catani2009entropy,greif2015formation, Mazurenko2017, Kantian2018, Chiu2018}, using the system's own degrees of freedom as a bath \cite{Paiva2011, Mathy2012, hart2015observation}.  
Recently, this technique has enabled the experimental observation of long-range antiferromagnetic order in quantum simulations of the Fermi-Hubbard model; in particular, two identical systems with extremely different densities were placed in contact with one another~\cite{Mazurenko2017,Chiu2018},  resulting in the emergence of an ultra-low entropy region~\cite{Kantian2018}.

\begin{figure}[t]
\includegraphics[width=0.9\linewidth]{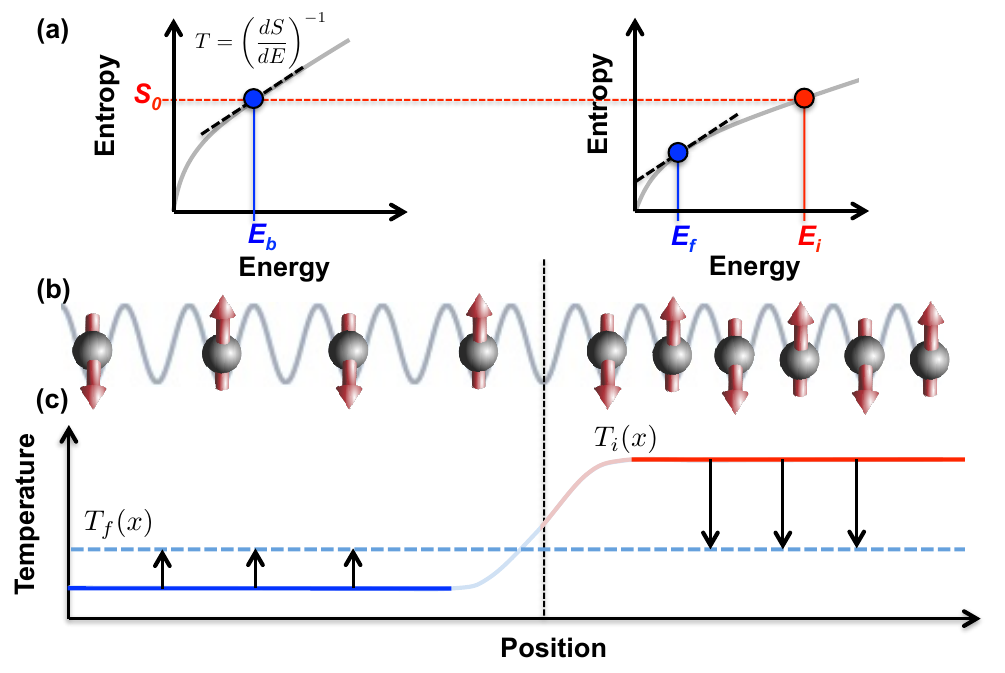}
\caption{a) If the Hamiltonian in the ``bath'' region is related to the Hamiltonian in the ``system'' region by a constant rescaling, $H_B = \lambda H_S$, their entropy-energy-density curves satisfy $s_B(E) = s_S(E/\lambda)$. Thus preparing at state with constant entropy density establishes a temperature differential $T_B = \lambda T_S$, since $T = \frac{d E}{d S}$.
b) Schematic representation of particles or spins interacting through a long-range, power-law interaction $1/R^\alpha$. If the interparticle spacing  on the left (``bath'') is increased by a factor $d$ relative to the right (``system''), then $H_B = (\frac{1}{d})^\alpha H_S$ \cite{moses2015creation}.
c) In this case, a uniform Ne\'el state has a temperature differential after reaching local equilibrium, and the resulting evolution will remove entropy from the right half of the chain as the system reaches global equilibrium.}
\label{fig:schematicquench}
\end{figure}

In this work, we propose and analyze a class of methods---termed `conformal cooling quenches'---for shifting entropy by spatially modulating the \emph{magnitude} of the Hamiltonian~\cite{magneticfridge}.
The intuition behind this approach is best illustrated as follows: Suppose that we take a system's Hamiltonian $H$ and either suddenly or adiabatically reduce it by a factor $\lambda < 1$, taking $H \to \lambda H$.
Since $k_B T$ has units of energy, the temperature $T$ is accordingly reduced by $T \to \lambda T$. When applied to the entire system, this ``cooling'' is trivial, since it amounts to a change of units without reducing the entropy density.
However, if the reduction by $\lambda$ instead occurs for only a \emph{portion} of the system, which we call the `bath,' the change in temperature is  physical, and establishes a temperature gradient; during equilibration, entropy will then flow out of the system and into the bath.

This generalizes previous studies, where entropy flow relies on particle itinerance, while the temperature gradient is inherited from a density gradient~\cite{Mazurenko2017, Kantian2018, Chiu2018}. 
In particular, our method is applicable not only to itinerant Hubbard systems, but also to spin models. This latter case is especially relevant to recent developments in trapped ion arrays~\cite{Britton2012, Zhang2017}, optical tweezer arrays~\cite{Labuhn2016, Bernien2017}, and ultracold molecules~\cite{Yan2013, Anderegg2019}, where versatile spin models with spatially tunable Hamiltonian parameters are increasingly accessible.

One virtue of the conformal cooling approach is that it can ``cool'' a system within a metastable state-space.  For example, conformal cooling can be applied to a gas equilibrating at negative kinetic or spin temperature \cite{braun2013negative}, bringing the system toward zero temperature from below.  It can also be applied to gases equilibrating in high-energy manifolds of states, i.e.~in excited bands of an optical lattice  \cite{muller2007state,wirth2011evidence}.  Systems equilibrating at negative temperatures or in higher bands can exhibit strong frustration without complicated band engineering.

We will begin by introducing the thermodynamics of our approach, focusing on two questions: 1) how much entropy can a cooling quench remove and 2) how long does it take? 
Next, we perform a large-scale numerical study of both a 1D topological spin-chain and the fermionic Hubbard model,  demonstrating that realistic cooling quenches can remove enough entropy to reveal their low-temperature physics. 
Finally, we discuss natural experimental implementations of our approach focusing on ultracold polar molecules and Rydberg atom arrays.

\emph{General Strategy}---We envision spatially demarcating the degrees of freedom  into a ``bath'' (B) and ``system'' (S) which are placed ``end-to-end,'' so that the coupling between their boundaries scales sub-extensively with their volume  (Fig.~\ref{fig:schematicquench}b)~\cite{sidebyside}. 
We assume  that the Hamiltonian $H_B$ (bath) is identical to $H_S$ (system), except that its magnitude is scaled  by a factor $\lambda < 1$. 
The entropy ($s$) versus energy density ($E$) curves in the two regions are then related by  $s_B(E) = s_S(E/\lambda)$ and their temperatures by $T_B(E) = \lambda T_S(E/\lambda)$ (Fig.~\ref{fig:schematicquench}a). 
In the following, we will consider two protocols, ``quenched'' and ``adiabatic.''

\emph{Quench Protocol}---In the quench approach, the Hamiltonians are time-independent with $H_B = \lambda H_S$. At $t=0$, we prepare  a uniform  initial state (e.g.~a product state) and simply let it evolve. Equivalently, one can begin in thermal equilibrium with $H_B =H_S$, and then suddenly reduce $H_B$ to $H_B = \lambda H_S$.  The overall system is now in \emph{local} equilibrium, with the local density matrices in $B$ and $S$ identical, and thus, $s_B = s_S$ and $T_B = \lambda T_S$.  As the system evolves toward \emph{global} equilibrium, entropy will follow the thermal gradient and  flow from $S$ to $B$.
%
%

\begin{figure}[t]
\includegraphics[width=0.7\linewidth]{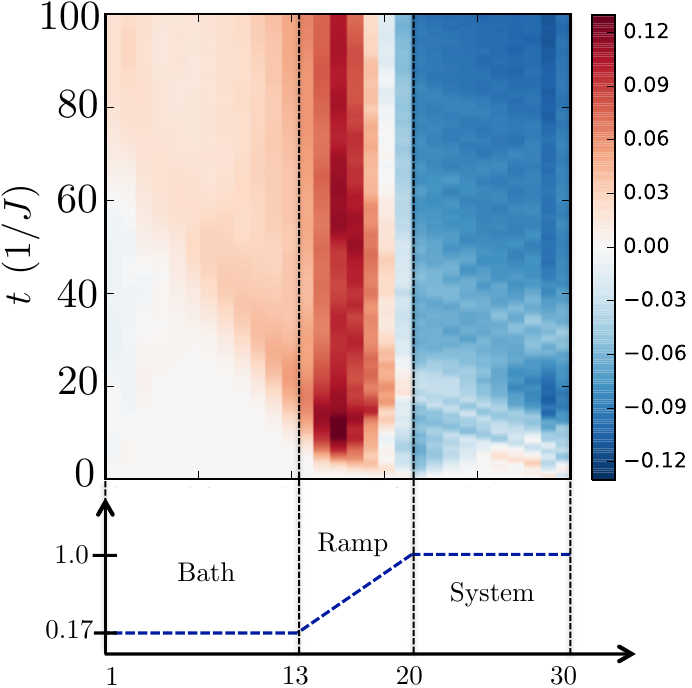}
\caption{ 
`Quench' cooling of a 30-site spin-1 Haldane chain. After initializing the state at $t=0$ with uniform entropy density, the coupling constants $\lambda_x$ are scaled according to the bottom panel, which should transport heat from the `system' on the right to the `bath' on the left. In the top panel, we plot the change in energy density $h_x(t) - h_x(0)$ as the chain evolves.
 }
\label{fig:aklt}
\end{figure}

The final equilibrium temperature $T_{\textrm{f}}^{(q)}$ is determined by energy conservation post-quench. Noting that the energy just after the quench is $N_S E_S(T_i) + N_B \lambda E_S(T_i)$, and using the relation $E_B(T) = \lambda E_S(T / \lambda)$, we have:
\begin{align}
 (N_S + \lambda N_B) E_S(T_{\textrm{i}}) = N_S E_S(T_{\textrm{f}}^{(q)}) + N_B \lambda E_S(T_{\textrm{f}}^{(q)}/ \lambda),
\label{eq:quench_TF}
\end{align}
where $N_S , N_B$ are the number of sites in the system and bath, and $T_{\textrm{i}}$ is the initial temperature of the system.
When $ \lambda N_B \gg N_S$, we have $T_{\textrm{f}}^{(q)} = \lambda T_i $, but more generally one should choose $\lambda$ so as to minimize $T_{\textrm{f}}^{(q)}$ based on the precise form of $E_S(T)$.
While we have assumed a sharp distinction between $S$ and $B$ for simplicity, one can let the spatial modulation $\lambda(\vec{x})$ vary smoothly, for example in the ``ramp'' region shown in Figs.~\ref{fig:aklt} and \ref{fig:hub_quench}(a), in which case Eq.~\eqref{eq:quench_TF} is replaced by an integral over the energy density.

\emph{Adiabatic Protocol}---The cooling is more effective if the magnitude of $H_B = \Lambda(t) H_S$ is instead slowly reduced in time, with $\Lambda(t=0)=1$ and $\Lambda(t \rightarrow \infty) = \lambda$.  In the isentropic limit, the final system temperature $T_{\textrm{f}}^{(a)}$ is determined by
\begin{align}
 (N_B + N_S) s_S(T_{\textrm{i}}) = N_B s_S(T_{\textrm{f}}^{(a)} / \lambda ) + N_S s_S(T_{\textrm{f}}^{(a)}),
\label{eq:adiabatic_cond}
\end{align}
with $T_{\textrm{f}}^{(a)}  \le T_{\textrm{f}}^{(q)}$.
When the bath and system are end-to-end, diffusive dynamics imply that the equilibration time, $t_{\textrm{eq}}$, scales as $L_S^{2} / \Lambda(t)$ (q.v. Eq.~\eqref{eq:timescale}) where $L_S$ is the linear extent of the system and adiabaticity requires $\partial_t \Lambda \ll 1/t_{\textrm{eq}}$. For small $\Lambda$, the bath and system will eventually fall out of equilibrium and additional entropy will be produced, though the temperature will always be upper-bounded by $T_{\textrm{f}}^{(q)}$. 

To demonstrate that conformal cooling can shift significant entropy out of the system even for moderate bath sizes and short time-scales, we numerically investigate two distinct settings: the $S=1$ Haldane topological anti-ferromagnet and the fermionic Hubbard model.

\emph{Conformal cooling in an $S=1$ Haldane chain}---Consider a one dimensional chain of $S=1$ spins with Hamiltonian 
\begin{align}
H[\lambda_x] = \sum_x \lambda_x h_x = J \sum_x \lambda_x \left[ \mathbf{S}_x \cdot  \mathbf{S}_{x+1}  + \frac{\gamma}{3} (\mathbf{S}_x \cdot  \mathbf{S}_{x+1})^2 \right].
\label{eq:haldane_chain}
\end{align}
At both the Heisenberg point $\gamma = 0$ and the AKLT point $\gamma  = 1$, the spin-chain is a gapped topological paramagnet in the  Haldane phase \cite{Haldane1983, AKLT1987}.
The topology of the phase has a striking signature in a finite-length chain, 
which exhibits a pair of localized spin-1/2 edge states.
At temperatures below the bulk gap, $T < \Delta \sim J$, these localized edge states can coherently store quantum information for long times,  providing a sharp experimental signature of the topological phase \cite{SenkoMonroeS1}. 

\begin{figure}[t]
\includegraphics[width=0.8\linewidth]{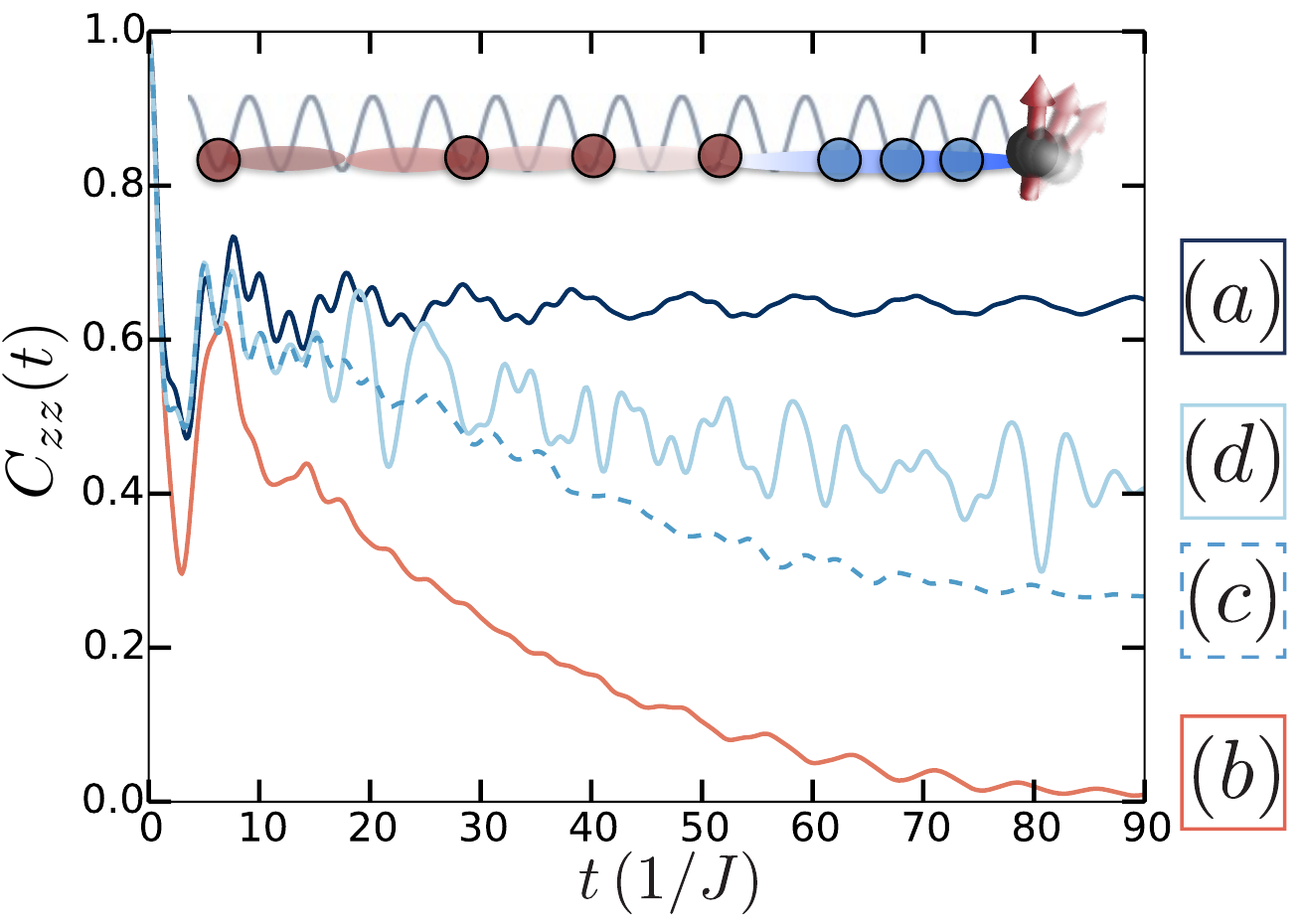}
\caption{The dynamical correlation function,  $C_{zz}(t)$, measured after four initialization protocols: a) Optimal: $C_{zz}(t)$ for the ground state of an $L=10$ chain; b) finite $T=0.51J$, without a cooling quench. The edge coherence rapidly decays. c) finite $T_i = 0.51J$, but starting the $C_{zz}$ measurement after the cooling quench shown in Fig.~\ref{fig:aklt}. The coherence is improved by an order of magnitude. d) Same as (c), but eliminating the coupling between sites $i=20$ and $21$ after the quench, which cuts off the bath. }
\label{fig:aklt_coherehence}
\end{figure}

Calculating the thermodynamic energy-temperature relation, $E(T)$, using exact diagonalization reveals that a modest bath size of $N_B / N_S\approx$ 2-3 is sufficient to cool from the Ne\'el  product state $\ket{\uparrow \downarrow \uparrow \downarrow \cdots}$, which corresponds to an initial temperature $T_{\textrm{i}} = 1.45 \Delta$, to well below the gap, $T_{\textrm{f}}^{(q)} \approx 0.7 \Delta$ \cite{suppinfo}. Here the pure-state temperature is defined by inverting $E(T)$.
Since the spin chain is diffusive \cite{DamleSachdev}, the time-scale required for  cooling  is determined by Fourier's law. When $\lambda(x)$ varies smoothly compared to the lattice scale,  the local thermal conductivity $\kappa$ and specific heat $c$ are determined by  rescaling, $\kappa(T, x) = \lambda(x) \kappa_S( T(x) / \lambda(x))$ and $c(T, x) =  c_S( T(x) / \lambda(x))$, where $\kappa_S(T)$ and $c_S(T)$ are defined with $\lambda = 1$.
Applying this within a simple lumped element model predicts that temperature will decrease as \cite{suppinfo}, 
\begin{align}
T_S(t) \sim T_{\textrm{f}}^{(q)}  + \mathcal{K} \frac{ L_S}{ \sqrt{t D_B}} (T_{\textrm{i}} - T_{\textrm{f}}^{(q)}) 
\label{eq:timescale}
\end{align}
where  $D_B$ is the thermal diffusivity of the bath  and $\mathcal{K}$ is an $\mathcal{O}(1)$ geometrical factor. 
For bath temperatures above $\lambda J$, the diffusivity will generically saturate to a temperature-independent value, $D_B \propto  \lambda J  / \hbar$  \cite{karadamoglou2004diffusive}, implying that  $t_{\textrm{eq}} \sim L_S^2 (\lambda J / \hbar)^{-1}$. 

To verify these dynamics, we simulate the evolution of a finite-energy density pure state using the TEBD-algorithm \cite{Vidal2003}.
It is exponentially difficult to simulate finite temperature dynamics, limiting our system to $L= 30$ sites (Fig.~2)~\cite{gamma34}. 
We initialize a uniform state $\ket{\Psi(0)} = e^{-\tau \hat{H}[\lambda_x = 1] } \ket{\uparrow \downarrow  \uparrow \downarrow\cdots}$, where  $\tau = 0.35 / J$, resulting in an energy density  that corresponds to temperature $T_S = 0.51 J $ after local equilibration~\cite{contrivedstate}.
The system is then quenched into a spatially non-uniform $\hat{H}[\lambda_x]$ (Fig.~2). Using the optimal $\lambda_0 = 0.17$ in the `bath' leads to a  final predicted temperature: $T_{\textrm{f}}^{(q)} = 0.29 J$.

The evolution of the local energy density  $\langle \lambda_x \hat{h}_x(t) \rangle$ during the cooling quench is depicted in Fig.~\ref{fig:aklt}.
The  energy density in region $S$ at time $t = 100/J$ corresponds to $T_{S} = 0.34J$, within $14\%$ of the expected $T_{\textrm{f}}^{(q)}$ \cite{suppinfo}.
Moreover, the relaxation dynamics are roughly consistent with $T_S \sim T_{\textrm{f}}^{(q)}+  (T_{\textrm{i}} - T_{\textrm{f}}^{(q)})  \sqrt{t_{\textrm{eq}} / t}$, where $t_{\textrm{eq}} \approx 0.22 ( \mathcal{K} L_S)^2 (\lambda J / \hbar)^{-1}$, consistent with the expectation $1/D_B  \sim 0.19 / \lambda$ \cite{karadamoglou2004diffusive, suppinfo}. 

Even for a relatively small bath size, the cooling quench has a dramatic effect on the dynamical  correlation function of the topological edge mode. 
Since the  edge state in region $S$ will generically have  overlap with the right-most spin $S^{\mu}_{\textrm{end}}$, its coherence can be probed via the correlation function
\begin{align}
C_{zz}(t) = \bra{\Psi}S_{\textrm{end}}^z(t + t_f) S_{\textrm{end}}^z(t_f) \ket{\Psi}, \end{align}
where the measurement only begins after the cooling quench is complete ($t_f = 100/J$).  
At $T=0$, these correlations should asymptote to a finite constant [Fig.~3a],
while at large $T$ [Fig.~3d], they will  decay exponentially.
We compare $C_{zz}(t)$ under four preparation scenarios described in Fig.~\ref{fig:aklt_coherehence}. The conformal cooling quench improves the coherence time (i.e.~the decay timescale of $C_{zz}(t)$) by more than an order of magnitude. 
	
\emph{Adiabatic conformal cooling in the fermionic Hubbard model}---We next consider the adiabatic protocol  applied to the fermionic Hubbard model, $H = -  \sum_{<i, j>, \sigma } t_{ij} c^\dagger_{i \sigma} c_{j \sigma} +  U \sum_i n_{i \uparrow} n_{i \downarrow} - \mu \sum_{i \sigma}  n_{i \sigma}$.
 Here, we focus on the   Mott insulating phase at half-filling with   $t / U \ll 1$ and $T < U$.
While the fermions' motion is quenched, their spins interact via an effective anti-ferromagnetic Heisenberg interaction, $H_{\textrm{eff}} =  \sum_{\left<i, j \right>} J_{ij} [ \mathbf{S}_i \cdot \mathbf{S}_j - \frac{1}{4} ]$,
where  $J = 4 \frac{t_{ij}^2}{U}$ is the super-exchange coupling.  

\begin{figure}[t]
\includegraphics[width=0.9\linewidth]{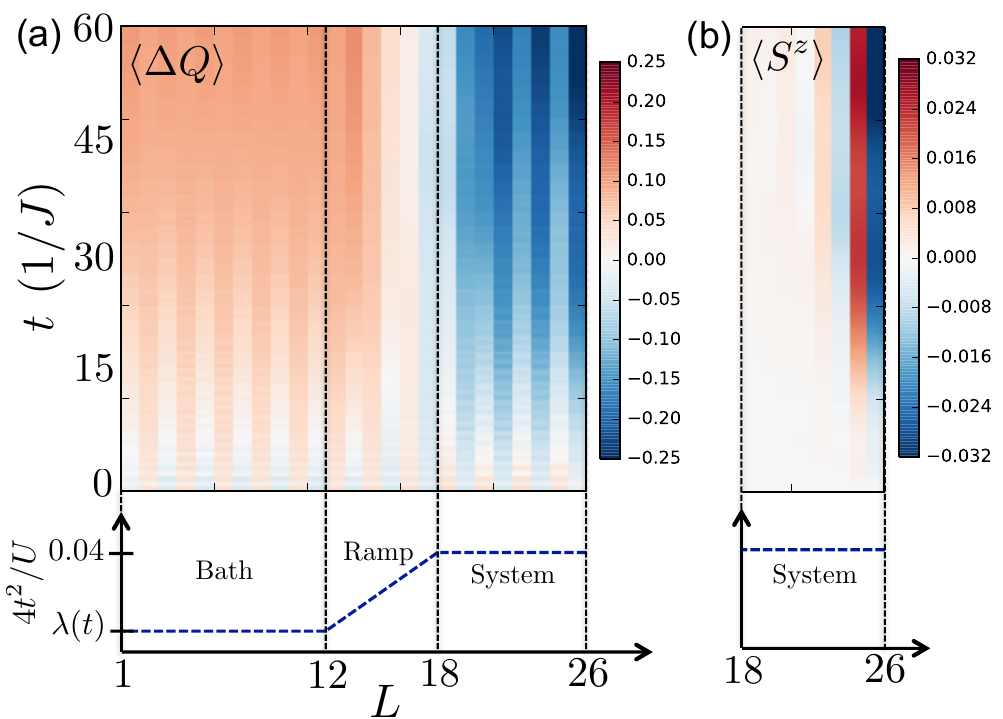}
\caption{ Cooling dynamics in the 1D spinful Hubbard model. a) After initializing a $T=1.4 J$ thermal state with uniform Hamiltonian $U=1, t_{ij} = 0.1$, the hopping $t_{ij}$ is adiabatically decreased with a spatial profile shown in the bottom panel. The top panel shows the change in the heat density as a function of time. b) Depicts the onset of antiferromagnetic correlations. The right-most site has a small Zeeman field $0.05 S^z$. While the initial temperature disorders the spins, as the system cools,  algebraic anti-ferromagnetic correlations clearly emerge.}
\label{fig:hub_quench}
\end{figure}

In the Mott regime where the dynamics are governed by $H_{\textrm{eff}}$,  adiabatic cooling is naturally realized by decreasing $J$ in the bath region (relative to the system region); one can  achieve this  by  weakly modulating the depth of the optical potential, $V(\vec{x}) = -V_0(\vec{x}) \sum_{i=1}^d \cos(k \, x^i)^2 $, where $V_0(\vec{x})$ is slowly varying and $k$ is the wavevector of optical lattice.
Increasing $V_0$ has three effects on the effective Hamiltonian: $U$ will increase, as the orbitals are further localized, $\mu$ will  increase, as the trap is deeper and $t$ will decrease due to the barrier height. 
Since $t$ is exponentially more sensitive than $U$ to the trap-depth, $\xi =\sqrt{ \frac{V_0}{E_r}}$ ($E_r$ is the recoil energy), the dominant effect is to modulate the hoppings \cite{hofstetter2002high}.
Assuming $\mu$ is  compensated  to maintain half-filling, the super-exchange energy becomes  $J_{ij} \propto \xi^2(x) e^{- 4 \xi(x) }$, precisely the desired modulation.
Fortuitously, a small modulation in $V_0$ is already capable of dramatically reducing the system's temperature; for example, in the the 3D cubic Heisenberg model, a $6\%$ change in the lattice depth  can cool the system from $1.4 T_{\textrm{N}}$ \cite{Hulet2015} down to the N\'eel temperature, $T_{\textrm{N}}$ \cite{suppinfo}.

Note that in the above approach, we choose to scale $t$ but not $U$, which  differs from the  overall scaling, $H \to \lambda H$, we had initially used to motivate our work.  Of course in the limit $t, T \ll U$, the conclusions are  the same  because the thermodynamics  are governed by $H_{\textrm{eff}} \propto J = 4 t^2 / U $, so scaling $t$ does effectively  enact an overall scaling of the Hamiltonian.
But more generally,  cooling only requires  the  criteria $\partial_t T(s; t,U) > 0$, which we have verified using determinantal quantum Monte Carlo \cite{suppinfo}, so long as the initial entropy density satisfies $s < k_B \log(2)$ \cite{dare2007interaction}. To this end, our proposal will also work away from the $t, T \ll U$ limit.

To confirm the effectiveness of the adiabatic protocol, we  simulate the dynamics of the spinful 1D fermionic Hubbard model. 
We use the TEBD method to time evolve a purified finite-temperature ensemble \cite{Karrasch2013}. At time $t=0$ the Hamiltonian is uniform, $U = 1, t_{ij} = 0.1$, with an initial thermal state  $\rho = e^{-H / T_i} / \mathcal{Z}$ at $T_i = 1.4 J$. 
We then time evolve the ensemble with a Hamiltonian, $H(t)$,  in which $t_{ij}$  decreases adiabatically in the bath \cite{suppinfo}.
Since Hamiltonian changes in time, energy is not conserved, and we divide it into heat and work, $\dot{E} = \dot{Q} - \dot{W}$ \cite{suppinfo}, enabling us to plot the evolution of the heat-density $Q$ in Fig.~\ref{fig:hub_quench}(a). Total heat is conserved, but with  clear transport from $S$ to $B$.
As a more qualitative thermometer, we note that at $T=0$, the system should display algebraic anti-ferromagnetic correlations. To reveal them, we place a  small Zeeman field $H = 0.05 S^z$ on the right edge spin, both in the initial thermal state and the subsequent dynamics. As depicted in Fig.~\ref{fig:hub_quench}(b), the finite temperature of the initial thermal state disorders the magnetization $\langle S^z\rangle$, but as the dynamics proceed and cooling occurs, the antiferromagnetic correlations become clearly manifest.

\emph{Experimental implementation}---Our conformal cooling protocols are  well suited to systems with long-ranged interactions, such as  polar molecules, Rydberg atoms, and trapped ions \cite{yan2013observation,moses2015creation,zeiher2016many,smith2015many,martinez2016real}. To implement the  quench protocol, we envision a setup where the  average spacing  between particles is larger in the bath than in the system, $r_B > r_S$.
Assuming power-law interactions ($1/R^\alpha$), the Hamiltonian in $B$ will be reduced by a constant factor $\lambda = (r_S / r_B)^\alpha$ relative to that in $S$ (Fig.~\ref{fig:schematicquench}b)~\cite{longrange}.

This approach is particularly applicable to two classes of current generation experimental platforms: ultracold polar molecules and Rydberg atom arrays. 
In the molecular context, the optical lattice filling fraction, $\nu < 1$, leads to random dilution \cite{moses2015creation}.
Fortunately, the cooling quench is natural to implement in this randomly diluted case, since one can make $r_B > r_S$ merely by modulating the average density, without having to ensure the particles in $B$ lie on a particular sub-lattice.
In this case, simply time-evolving an initial product state in the presence of this density modulation will cool the high-density region, and could provide a simple route towards studying, for example, algebraically correlated random-singlet phases \cite{gorshkov2011tunable, yao2015quantum, fisher1994random}.

Although we have studied the AKLT model because it admits simple numerical observables for quantifying entropy, the same cooling protocol can also be applied to the long-range, mixed-field Ising model, which is naturally realized in a Rydberg atom array~\cite{Labuhn2016, Bernien2017,keesling2019quantum, Lahaye2019, Browaeys2020}. 
In this case, the spacing between the atoms and/or the intensity of the Rydberg excitation light, can be made spatially varying, in order to create well-defined bath and system regions in one, two, or even three-dimensions~\cite{endres2016, Barredo2016, Barredo2018}.
The complex phase diagram associated with this model exhibits a variety of competing orders and phase transitions, providing a rich playground for implementing conformal cooling \cite{samajdar2020complex}. 
%

In summary, we have proposed a general method for preparing low-entropy many-body states in isolated quantum systems. Our approach can be naturally implemented in systems with power law interactions by simply diluting the particle density of the bath region; moreover, in the supplemental materials, we also provide a simple experimental blueprint for implementing conformal cooling in the spinful fermionic Hubbard model \cite{suppinfo}. Looking forward, our proposal raises a number of intriguing questions: is it possible to  implement a refrigeration cycle by repeated preparation of the bath state? Can one optimize a side-by-side geometry which could reduce the equilibration time?  By performing conformal cooling during a quantum phase transition, can one reduce the rate of Kibble-Zurek defect formation \cite{keesling2019quantum}?

\begin{acknowledgments}
	We thank Randy Hulet, Jun Ye, and Martin Zwierlein for insightful suggestions and illuminating conversations. We acknowledge the QUEST-DQMC collaboration \cite{QUEST} for providing the code used in our QMC calculations. This work was supported by the ARO through the MURI program (W911NF-17-1-0323 and W911NF-20-1-0136), the President's Research Catalyst Award CA-15-327861 from the University of California Office of the President, the David and Lucile Packard foundation and the W.~M.~Keck foundation. A.~M.~K. acknowledges support from NIST.
\end{acknowledgments} 

\bibliography{cooling}

\begin{thebibliography}{73}
\bibitem [{sid()}]{sidebyside}%
  \BibitemOpen
  \href@noop {} {}\bibinfo {note} {In 1D and 2D, it may be possible to place
  $S$ and $B$ ``side-by-side,'' so that their contact scales extensively with
  their volume. This arrangement could reduce the equilibration time, since one
  no longer requires diffusion. However, the coupling between $S$ and $B$ now
  has a non-trivial contribution to the thermodynamics and its effect would
  have to be accounted for in detail.}\BibitemShut {Stop}%
\bibitem [{gam()}]{gamma34}%
  \BibitemOpen
  \href@noop {} {}\bibinfo {note} {We work at $\gamma = \frac{3}{4}$, which
  reduces the correlation length relative to $\gamma=0$ but is otherwise
  completely generic.}\BibitemShut {Stop}%
\bibitem [{con()}]{contrivedstate}%
  \BibitemOpen
  \href@noop {} {}\bibinfo {note} {Note that while this state seems contrived
  from an experimental perspective, it is convenient to prepare numerically,
  and owing to ergodicity the only relevant characteristic is its energy
  density.}\BibitemShut {Stop}%
\bibitem [{lon()}]{longrange}%
  \BibitemOpen
  \href@noop {} {}\bibinfo {note} {We note that in our TEBD studies, we have
  focused on models with short-range interactions owing to numerical
  tractability. One might wonder whether long-range interactions still allow
  for a clear distinction between the system and bath regions. Crucially, the
  ability to distinguish between the system and bath in the thermodynamic limit
  does not require a strictly short-range interaction, only that the
  interfacial contribution to the free energy scales with the interfaces' area
  rather than with the volume of the bulk. For example, for a $1/R^\alpha$
  power-law interactions in $d$-dimensions, degrees of freedom interact with a
  net strength of $R^{d-1 - \alpha}$ with degrees of freedom at distance $R$.
  Thus, this condition is naturally satisfied if $\alpha + 1 > d$, which is the
  case for both dipolar molecules ($\alpha = 3$) and Rydberg arrays ($\alpha =
  6$).}\BibitemShut {Stop}%
\end{thebibliography}

\onecolumngrid
\appendix

\end{document}


\title{Supplemental Material for Preparation of Low Entropy Correlated Many-body States via Conformal Cooling Quenches}
\author{M. P. Zaletel, A. M. Kaufman, D. M. Stamper-Kurn, N. Y. Yao}
\maketitle

\section{Thermodynamics of the quench protocol for the $S=1$ chain}
\begin{figure}[h]
\includegraphics[width=0.7\linewidth]{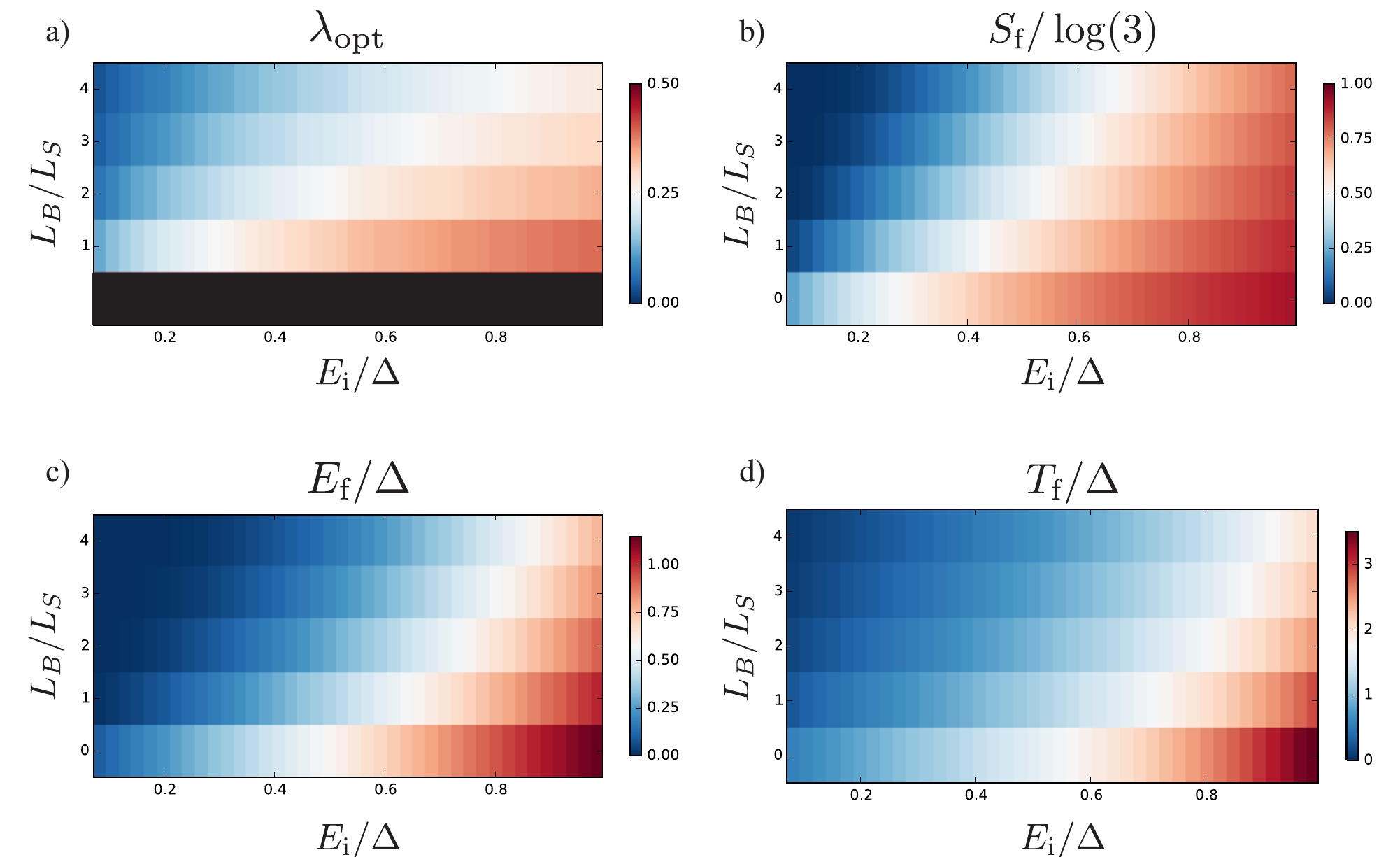}
\caption{Thermodynamics of the cooling quench for a spin-1 Heisenberg chain at $\gamma = 3/4$ (results for other $\gamma$ are similar).
The final state of the system depends on the ratio of bath to system size, $L_B / L_S$, and the initial energy density above the ground state in unit of the gap, $E_\textrm{i} / \Delta$. For comparison, the state $\ket{\uparrow \downarrow \cdots}$ is at $E_\textrm{i}/ \Delta = 0.54$.
(\textbf{a}) The optimal choice of bath scaling $\lambda_{\textrm{opt}}$. Larger baths allow for smaller $\lambda_{\textrm{opt}}$, and hence greater cooling.
(\textbf{b}) The final system entropy $S_\textrm{f}$ in units of the infinite-temperature value $\log(3) k_B$. (\textbf{c}) The final system energy density above the  ground state, $E_{\textrm{f}}$, in units of the gap. (\textbf{d}) 
The final system temperature, $T_{\textrm{f}}$, in units of the gap.}
\label{fig:quench_table}
\end{figure}

Here we report the final system energy density, entropy density, and temperature  as a function of both the initial energy density in units of the gap, $E_\textrm{i} / \Delta$, and the ratio of bath to system length $L_B / L_S$.
The thermodynamic relation $T(E)$ was computed using exact diagonalization Eq.~(3) for 10-site chain, which we then use to numerically solve Eq.~(1). A table of results is reported in Fig.~\ref{fig:quench_table}.

\vspace{3mm}
\section{Diffusion times for the cooling quench}
\begin{figure}[t]
\includegraphics[width=0.4\linewidth]{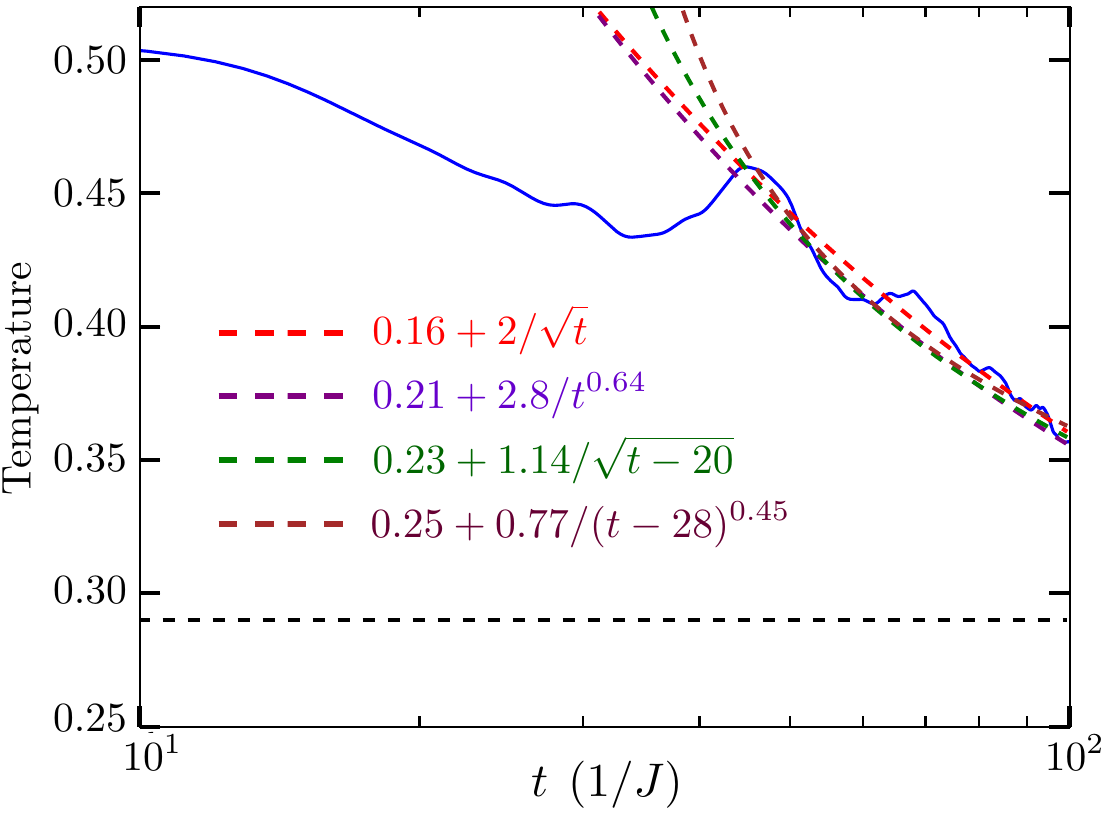}
\caption{The system temperature $T_S(t)$ at time $t$ of the Haldane-chain cooling quench shown in Fig.~2 of the main text. $T_S$ is inferred by measuring the energy density at the center of $S$, using the thermodynamic relation $T_S(E)$ computed from exact diagonalization.
Unfortunately we can only evolve out to about $t = 100/J$ (which takes several weeks of computing time), so it is difficult to tell how well the decay matches the expected $\sqrt{t_{\textrm{eq}}/t}$ behavior.
The predicted final temperature is shown as a horizontal dashed line, though a deviation due to finite size effects would be expected for such a small system. Various ansatz fits our shown, which show rough consistency with the expected $1/\sqrt{t}$ behaviour.
}
\label{fig:sqrttime}
\end{figure}

Solving the heat diffusion equation exactly is not possible, since the temperature dependence of the heat capacity and thermal conductivity leads to a non-linear PDE that depends on the details of the model.
As an illustrative approximation, note that temperature gradients in $S$ decay exponentially over the diffusion time $ t_S \sim L_S^2 / D_S$,
while when $V_B$ is large the average temperature $T_S$ of the system  decays as a much slower power law, $T_S \sim 1 / \sqrt{t}$.
Since the dynamics of $T_S$ are so much slower than the dynamics of the temperature gradients in $S$, we can consider a simplified model which lumps the system together into a single heat capacity $C_S(T_S) = V_S c_S(T_S)$.
As for the bath, we assume it is at a relatively high temperature in terms of the natural units $\lambda J$. In a high-temperature expansion for a spin model, both the heat capacity and thermal conductance scale as $1/T^2$, so the diffusivity $D_B  = \kappa / c$ is nearly constant ( $D_B = 5.3 \lambda J  / \hbar$ for the AKLT chain \cite{Karadamoglou2004diffusive}).
Thus we consider a continuous medium of diffusivity $D_B$ coupled to a single heat capacity $C_S$, where at time $t = 0$ the system temperature is $T_S = T_i$ and the bath temperature is $T_B = \lambda T_i$.
Ignoring  the temperature dependence of $c_S / c_B$, Fourier's law can be solved to obtain
\begin{align}
T_S(t) \sim T_{\textrm{f}}  + \mathcal{K} \frac{c_S}{c_B} \frac{ L_S}{ \sqrt{t D_B}} (T_{\textrm{i}} - T_{\textrm{f}}).
\label{eq:timescale}
\end{align}
where $\mathcal{K}$ is an order-one geometrical factor and $T_{\textrm{f}}$ is the final equilibrium temperature.
After the system reaches $T_S < \Delta$, with $\Delta$ the gap, $c_S / c_B$ will decrease exponentially with $T_S$ (since the system's heat capacity is activated), so ignoring the  temperature dependence of $c_S / c_B$ is no longer justified.
This does not accelerate the cooling, however; the $t^{-1/2}$ behavior will remain since the temperature gradient in the \emph{bath} must still relax. In Fig.~\ref{fig:sqrttime}, we show $T_S(t)$ for the cooling quench of the Haldane model.

%

\section{Definition of $Q$ for adiabatic quench.}

	In the adiabatic protocol the Hamiltonian $H(t) = \sum_i h_i(t)$ changes in time, so energy is not conserved. Nevertheless, we would like a convenient way to show heat is being transported between the bath and system. For a large system we could  use the local temperature as inferred from the local energy density, but for the small chains studied here we find  artifacts near the edge and ramp which make it less useful to visualize.
	
\begin{figure}[h]
\includegraphics[width=0.7\linewidth]{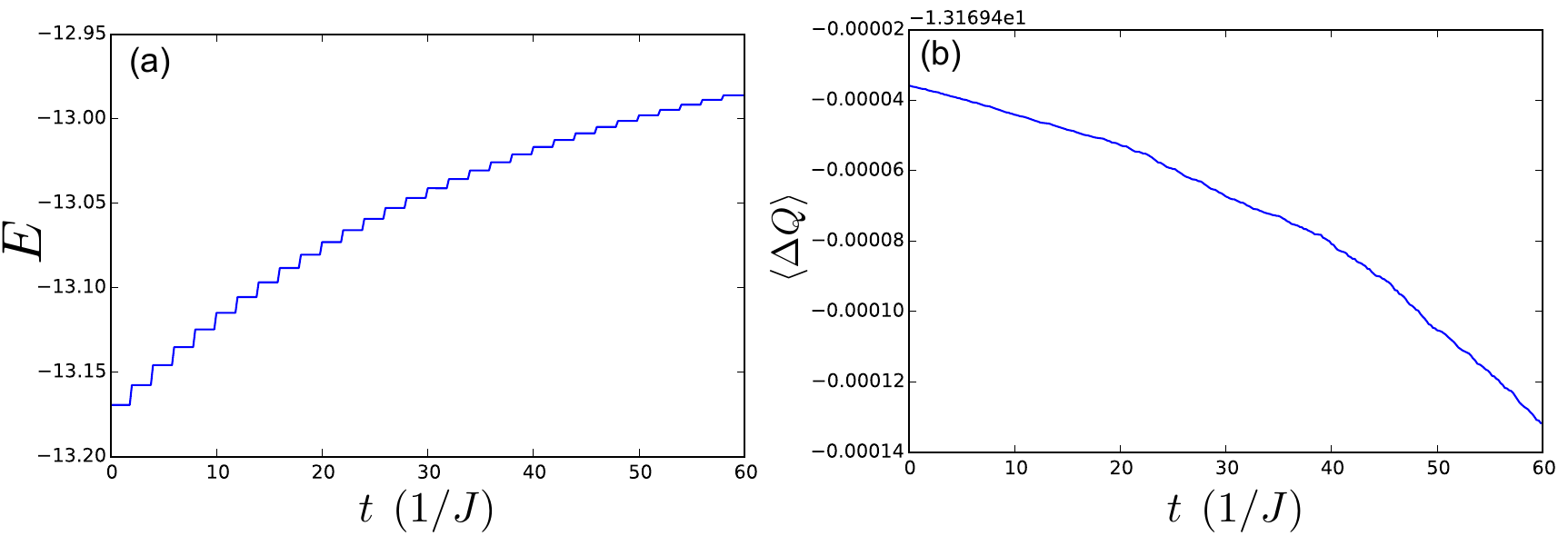}
\caption{(a) Total energy $E$ as a function of time for the 1D Hubbard simulations. For numerical convenience, the coupling constants are changed in a series of small, discrete steps, leading to step-discontinuities in the energy interlaced with periods of $E$-conserving evolution. Likewise, the local energy density $E_i$ has step-discontinuities interlaced with slow dynamics. $Q_i$ is conveniently obtained from the $E_i$ by  dropping all the step-discontinuities.
(b) The total $Q$ is conserved to better than $10^{-4} J$; the error aries from numerical errors like the finite Trotter step used in the finite-temperature TEBD dynamics.
}
\label{fig:hubbardens}
\end{figure}

	Instead, we note that the energy changes according to $\partial_t E =  \sum_i \bra{t} (\partial_t h_i) \ket{t}$, and that the change in energy can be divided into the heat $Q$ and work $W$, $dE = \delta Q - \delta W$.
In the adiabatic limit $\delta Q = 0$ globally, but heat can be transported between regions (of course transport of energy is driven by entropy production, so this transport  itself must be arbitrarily slow in the adiabatic limit).
This motivates the definition $Q = \sum_i Q_i$ where $\partial_t Q_i =  \partial_t  ( \bra{t}  h_i \ket{t} )  - \bra{t} (\partial_t h_i) \ket{t}$.
By construction $\partial_t Q = 0$, though each $Q_i$ may change. It is these $Q_i$ which are shown in the Fig. 4 of the main text.
The practical implementation of this subtraction is illustrated in Fig.~\ref{fig:hubbardens}.

While the $Q_i$ can be defined in this manner away from the adiabatic limit, and $\sum_i Q_i$ is always constant, it does not have the precise definition of `heat,' since the diffusive transport itself generates entropy, and hence heat $T dS$. Note, however, that in the system $\partial_t h_i = 0$, so there $Q_i$ reduces to the energy density. Since our experiment shows $\partial_t Q_i < 0$ in the system, it is unambiguously cooling.

\section{Adiabatic cooling of Hubbard model outside the Mott regime}
The analysis of the Hubbard model in the main text assumed the Mott limit $t / U \to 0$, so that $t, U$ only appear through the super-exchange energy $J = 4 \frac{t^2}{U}$. To confirm the approach is effective outside the Mott regime, we note that for the adiabatic cooling to work we only require that the temperature satisfies $\partial_t T(s; t, U) > 0$, where $s$ is the entropy density.
Using determinantal quantum Monte Carlo to determine $T$ for the 2D Hubbard model, we see this condition it satisfied whenever $s < \log(2) k_B$~\cite{Tremblay2007}.

\begin{figure}[h]
\includegraphics[width=0.35\linewidth]{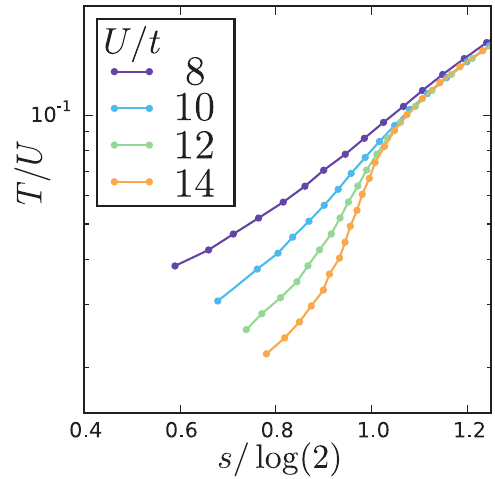}
\caption{Thermodynamics of the 2D fermionic Hubbard model: dependence of the temperature $T$ on the entropy density $s$ and the model parameters $t, U$, at half-filling. We see that at fixed entropy, $\partial_t T|_{s} >0$ whenever $s < \log(2) k_B$, implying the adiabatic scheme works outside the Mott limit. Calculations were done using the QUEST DQMC package on an $8 \times 8$ torus with extrapolation of the Trotter step to zero.}
\label{fig:dqmc}
\end{figure} 

\section{Experimental proposal for adiabatic conformal cooling in a cold-atomic fermi-Hubbard simulation}

In this section, we describe an experimental route toward realizing the adiabatic cooling protocol in a cold-atomic fermi-Hubbard simulation. As discussed in the main text, the key challenge is the generation of a modulated optical potential, $V_0(\vec{x})$, while keeping the chemical potential nearly constant. While the required change in lattice depth $V_0$ is modest, the modulated potential nonetheless requires the ability to create two differing lattice potentials in the system and bath regions. The intermediate transitional region does not need to be spatially sharp and a ``ramp'' region (main text, Fig.~4) will help to ensure that thermal diffusion remains efficient across this boundary. 

In principle, a quantum gas microscope combined with digital mirror devices or spatial light modulators allows one to paint a near-arbitrary optical potential.
However, since we require only the special case of a ``binary'' optical potential, we propose a simpler strategy based upon overlaying two distinct optical intensity patterns that are projected onto a single object plane.  
Rather than using a mirror for retro-reflection, each of the lattices $B$, $S$ can be generated by two incoming counter-propogating beams, for a total of four beams: $B_1, B_2, S_1, S_2$ (Fig.~S5).
By modulating the intensity of both $B_1$ and $B_2$, one can independently control the periodic component $V_0$ and the slow component $\mu$ (likewise for $S$).
Before reaching the object plane, we envision a binary spatial filter that transmits pattern $S_1, S_2$ to the system's spatial region and pattern $B_1, B_2$ to the bath's spatial region. 
Such spatial filtering can be easily achieved in two ways: 1) by utilizing a spatially patterned wave-plate followed by a polarizing filter or 2) by using a dichroic interference filter (Fig.~S5).

 \begin{figure}[t!]
\includegraphics[width=2.6in]{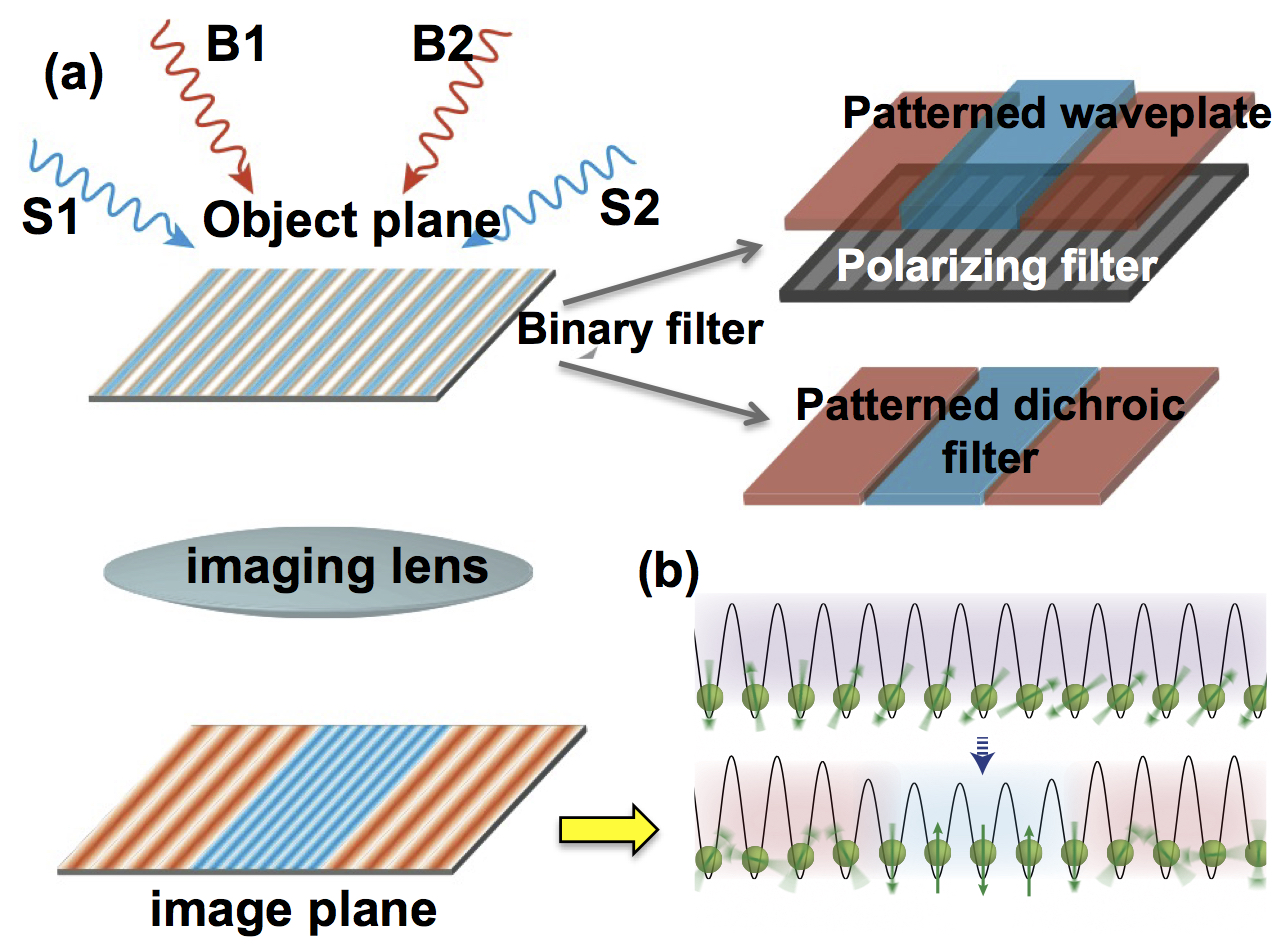}
\caption{a) Schematic of a ``binary'' optical potential based upon  overlaying two distinct optical intensity patterns that are projected onto a single object plane. Binary filtering can be achieved using either a spatially patterned wave-plate followed by a polarizing filter or via a dichroic interference filter.  b) The resultant binary potential  modulates the depth of the optical lattice yielding a super-exchange Hamiltonian with smaller overall magnitude in the bath.} 
\label{fig:hub}
\end{figure}

\bibliography{cooling}